\documentclass[a4paper,10pt]{article}
\usepackage{float,multicol,amsmath,amssymb,array,graphicx}
\def\JLone<#1,#2>{#1}
\def\JLtwo<#1,#2,#3>{#2}
\def\JLyear<#1,#2,#3,#4>{#3}
\def\JLpage<#1,#2,#3,#4>{#4}
\newcommand\JL[1]{\@JLone<#1>\ {\bfseries \JLtwo<#1>} (\JLyear<#1>), \JLpage<#1>}
\def\Jpage<#1,#2,#3>{#3}
\newcommand\andvol[1]{{\bfseries \JLone<#1>} (\JLtwo<#1>), \Jpage<#1>}
\newcommand\PTP[1]{Prog.\ Theor.\ Phys.\ \andvol{#1}}
\newcommand\PTPS[1]{Prog. Theor. Phys. Suppl.\ {No. \JLone<#1> (\JLtwo<#1>), \Jpage<#1>}}

\newcommand\PRD[1]{Phys.\ Rev.\ D\ \andvol{#1}}

\newcommand\PRL[1]{Phys.\ Rev.\ Lett.\ \andvol{#1}}

\newcommand\PLB[1]{Phys.\ Lett.\ B\ \andvol{#1}}

\newcommand\NPB[1]{Nucl.\ Phys.\ B\ \andvol{#1}}
\newcommand\PRP[1]{Phys. Rep.\ \andvol{#1}}
\newcommand\JHEP[1]{J. High Energy Phys.\ \andvol{#1}}
\begin{document}

\begin{center}
{\Large\bf  Dilatonic Inflation, Gravitino and Reheating
 \\ in Modified Modular invariant Supergravity} \\
\vspace{16pt}
Yuta Koshimizu$^{a}$, Toyokazu Fukuoka$^{a}$, Kenji Takagi$^{b}$, Hikoya Kasari$^{a}$\\
 and Mitsuo J. Hayashi$^{a}$ \\
\vspace{16pt}
$^{a}${\it Department of Physics, Tokai University, Hiratsuka, 259-1292, Japan} \\
$^{b}${\it Company VSN, Minato, Shibaura, 108-0023, Japan} \\

\vskip 16pt
E-mail: mhayashi@keyaki.cc.u-tokai.ac.jp
\vskip 16pt
\today
\end{center}

\thispagestyle{empty}
\vspace{24pt}

\begin{abstract}
A new modified string-inspired modular invariant supergravity model is proposed and is applied to realize the slow roll inflation in Einstein frame, so that the model explains WMAP observations very well.
Gravitino mass and their production rate from scalar fields are estimated at certain values of parameters in the model.
Seven cases of parameter choices are discussed here, among which some examples show the possibility of observation of gauginos by LHC experiments, which will give some hints of identity of dark matters.
The reheating temperature, which is estimated by the stability condition of Boltzmann equation by using the decay rates of the dilaton $S$ into gauginos, is lower than the mass of gravitino. Therefore no thermal reproduction of gravitinos happens. 
The ratio between the scalar and tensor power spectrum is predicted to be almost the same for the seven cases under study, and its value $r \sim 6.8 \times 10^{-2}$ seems in the range possibly observed by the Planck satellite soon. 
The plausible supergravity model of inflation, which will be described here, will open the hope to construct a realistic theory of particles and cosmology in this framework, including yet undetected objects. 
\end{abstract}

\newpage

\section{Introduction}

In order to construct an inflation model in supergravity (with slow roll conditions) successfully, one must at least explain observations appropriately \cite{Komatsu:2010fb,Jarosik:2010iu}. If this is the case, the model may have possibilities to predict objects of supergravity properly, such as gravitino mass, gaugino mass and the masses of superpartners of scalar fields. If the predicted values are experimentally testified, then the supergravity model has some plausibility as a true theory.
  In this paper we will propose a new modular invariant model, which satisfies the above requirement. 
It is convenient to introduce the dilaton superfield $S$, a chiral superfield $Y$ (gaugino condensate superfield), and the modular superfield $T$. Here, the other matter fields are set to zero for simplicity.
Moreover, we identify the inflaton field with the dilaton. Because the inflation concerns Planck scale physics, the dilaton seems the most appropriate candidate for the inflaton field. (Though we have exclusively restricted our attention to a model similar to Ref.\cite{02,03,04,05}, other models derived from another type of compactification seem very interesting. Among them, the KKLT model \cite{Kachru:2003aw, Endo:2005uy, Choi:2005uz, Choi:2005ge, Lebedev:2006qq, Lebedev:2006qc} attracts our interest, where the moduli of the superfield $T$ play essential roles.)
Seven cases of parameter choices will be discussed here, among which some examples show the possibility of observation of gauginos by LHC experiments, which will give some hints of identity of dark matters.\\
We first review the inflationary parameters obtained in the slow roll approximation (SRA) \cite{ref:Liddle,ref:Perturbation}.  
The slow-roll parameters (in Planck units $M_{\rm P}/\sqrt{8\pi}=1$) are defined by:
\begin{equation}
\epsilon_\alpha=\frac{1}{2}\left(\frac{\partial_\alpha V}{V}\right)^2, \quad\quad 
\eta_{\alpha\beta}=\frac{\partial_\alpha\partial_\beta V}{V}.
\end{equation}
The slow-roll condition is well satisfied, and the $\eta$-problem can just be resolved as shown below. 
Using SRA, number of $e$-folds $N$ at which a comoving scale $k$ crosses the Hubble scale $aH$ during inflation is given by:
\begin{equation}
N \sim -\int^{S_{\rm end}}_{S_*}\frac{V}{\partial V}dS,
\end{equation}
where the integral extends from $S_{\rm end}$ to $S_*$, where $S_{\rm end}$ is the value of $S$ at the end of inflation and $S_*$ is that at the beginning of inflation.
Next, a scalar spectral index $n_S$ for the scale dependence of the spectrum of density perturbation and its tilt $\alpha_S$ are defined by:
\begin{equation}
n_S - 1 = \frac{d\ln \mathcal{P_R}}{d\ln k}, \quad\quad
\alpha_s=\frac{dn_s}{d\ln k}.
\end{equation}
These are approximated in the slow-roll paradigm as:
\begin{equation}
n_S (S)\sim 1-6\epsilon_S+2\eta_{SS}, \quad\quad
\alpha_S (S)\sim 16\epsilon_S\eta_{SS} -24\epsilon_S^2-2\xi^2_{(3)},
\end{equation}
where $\xi_{(3)}$ is an extra slow-roll parameter that includes the third derivative of the potential as follows: 
\begin{equation}
\xi^2_{(3)} \equiv \frac{1}{64 \pi^2 G^2} \frac{V_S V_{SSS}}{V^2}.
\end{equation} 
Substituting $S_*$ into these expressions, we get
finally the following estimate for the spectrum of the density perturbation $\mathcal{P_R}$ caused by slow-rolling dilaton:
\begin{equation}
\mathcal{P_R} \sim \frac{1}{12\pi^2}\frac{V^3}{\partial V^2}.
\end{equation}
The spectrum of the density perturbation of tensor type $\mathcal{P}_T$ is given by 
\begin{equation}
\mathcal{P}_T = 64 \pi G \left( \frac{H}{2\pi} \right)^2_{k=aH} = \left( \frac{2}{3\pi^2} V \right)_{k=aH}.
\end{equation}
The spectral index $n_T$ is then given as
\begin{equation}
n_T = -2 \varepsilon. 
\end{equation}
The tilt is given by
\begin{equation}
\alpha_T=-4 \varepsilon ( 2 \varepsilon - \eta).
\end{equation}
Moreover, the ratio between scalar power spectrum and tensor one, which is  denoted by $r$, is also very important:
\begin{equation}
r \equiv\frac{\mathcal{P}_T}{\mathcal{P_R}} = 16 \varepsilon = - 8 n_T .
\end{equation}
The upper limit $r<0.24$ is given by seven years observation by WMAP. We hope that the Planck satellite might give the direct signal of primordial tensor perturbation soon. Or, if the observation of primordial non-Gaussianity indirectly proves the contribution of tensor mode from nonlinear terms of the perturbation spectrum, the value of the ratio will be improved. 
The current $68 \%$ limit from the 7-year data of WMAP is the primordial non-Gaussianity parameter $f_{NL} = 32 \pm 21$, and the Planck 
satellite is expected to reduce the uncertainty by a factor of four in a few years from now\cite{Komatsu, Tanaka}. We 
will show the prediction of the ratio corresponding to each parameter choice in Tables \ref{tab:infparam1} and \ref{tab:infparam2}.
Seven cases of parameter choices will be  discussed in this paper, which fit very well with the 7-year WMAP observations. 
It is the end of inflation, when the slow-roll parameter $\epsilon_\alpha$ or $\eta_{\alpha\beta}$ reaches the value 1. 
After passing through the minimum of the potential, reheating will begin. \\
In Section 2, a Modified String-inspired Modular invariant Supergravity is newly proposed.
In Section 3,  the Inflationary Cosmology and Inflationary Trajectory of this model are presented, in which the numerical values are explained by using the case 1 of parameters.
In Section 4, super Higgs mechanism and gravitino production from heavy scalar field are discussed.
In Section 5, reheating temperature is estimated by calculating the decay rate of inflaton into gauginos in minimal supersymmetric standard model (MSSM).
In Section 6, numerical predictions by tuning parameters are explained and shown in Tables \ref{tab:infparam1} and \ref{tab:infparam2}.
Finally, a short summary is given in Section 7. 

\section{ A Modified String-inspired Modular invariant Supergravity}

${\mathcal N} = 1$, $d = 4$ supergravity from $d = 10$ heterotic string by dimensional reduction has No-scale structure with $E_8 \times E_8$ gauge group \cite{01}. 
The modular invariant supergravity model was proposed, where the K\"{a}hler potential and the superpotential are given as:
\begin{eqnarray}
\!\!\!\!\!\!\!K=-\ln \!\left(S+S^\ast\right) -3\ln \!\left(T+T^\ast-|Y|^2 \right), \quad
W = 3bY^3\ln\left[c\>e^{S/3b}\>Y\eta^2(T)\right].
\end{eqnarray}
Here $\eta$ is  Dedekind's  $\eta$ function, $c$ is a free parameter of the theory, $S$ is the dilaton, $T$ is 
the moduli and $Y$ is a complex scalar superfield  defined by the gaugino condensation $U\sim <\lambda\lambda>=Y^3$ of the $E_8$ hidden sector\cite{02}.
The renormalization group parameter $b=\frac{15}{16\pi^2}$ can correspond to the $E_8$ hidden sector gauge group.

However, if we derive the scalar potential from these formulae in the Einstein frame, it seems impossible to solve difficulties such as the $\eta$-problem and/or the negative vacuum energy.
We would like to propose a new modular invariant  ${\mathcal N} = 1$ supergravity, which solve such problems.
A modified string-inspired modular invariant supergravity is proposed here to apply it to inflationary cosmology.\\ 
In the original model, the massless Goldstino was given by the the dilatino $\tilde{S}$ as shown in our former study, 
because $m_{SS}=0$,
where $m$ is defined\cite{ref:Kallosh} by $m \equiv e^{K/2} W$ and $G \equiv K+\log W^*W$ as
\begin{eqnarray}
m^{ij}=\mathcal{D}^i\mathcal{D}^j=[\partial^i+\frac{1}{2}(\partial^iK)]m^j-\Gamma_k^{ij}m^k,\qquad 
\Gamma_i^{jk}=G_{\ \ \ i}^{-1l}\partial^jG_l^k.
\end{eqnarray}. 
In order to extend the original model to realize the slow roll inflation with $m_{SS}=0$, the simplest choice is to add a term linear term in $S$, like $\alpha+\beta S$: 
\begin{eqnarray}
W=\alpha+\beta S+3bY^3\ln\left[c\>e^{S/3b}\>Y\eta^2(T)\right],
\end{eqnarray} 
where $\alpha$ and $\beta$ are new parameters that should be determined from observations. 
Then the scalar potential $V_E ( \equiv e^G \left[ G_i G^{ij^*} G_{j^*} - 3 \right] )$ is obtained as: \\[10pt]
\begin{eqnarray}
V_E &=& \frac{1}{(S+S^*)(T+T^*-|Y|^2)^2}  \left[ 3 b^2 |Y|^4 \left| 1 + 3 \ln \left[ c\>e^{S/3b}\>Y\eta^2(T) \right] \right|^2 \right. \nonumber \\[5pt]
&& + \frac{1}{T+T^*-|Y|^2} \left| \strut \alpha + \beta S + 3 b Y^3 \ln \left[ c\>e^{S/3b}\>Y\eta^2(T) \right] - (S+S^*) (Y^3 + \beta) \right|^2  \nonumber \\[5pt]
&& + 6 b^2 |Y|^6 \left\{ \left( 1- \frac{\alpha + \beta S^*}{b{Y^*}^3} \right) \frac{ \eta' (T)}{ \eta (T)} + \left( 1- \frac{\alpha + \beta S}{bY^3} \right) \frac{ \eta' (T^*)}{ \eta (T^*)}   \right.  \nonumber \\[5pt]
&&\qquad\qquad\qquad\qquad\qquad\qquad\qquad\qquad\qquad
 \left. \left. + 2 (T+T^*) \left| \frac{ \eta' (T)}{ \eta (T)} \right|^2 \right\} \right],
\end{eqnarray}
which is explicitly modular invariant in $T$. 
Instead of imposing $W_Y+K_YW=0$, we will assume $W_Y=0$, which is a rather good approximation, i.e.,
\begin{eqnarray}
W_Y = 3bY^2 + 9bY^2 \ln \left[ c \exp \left( \frac{S}{3b} \right) Y \eta^2 (T) \right]=0,
\end{eqnarray}
Then a relation between $S$ and $Y$ is obtained as follows:
\begin{eqnarray}
Y \sim \frac{1}{c \eta^2 (T) e^{\frac{1}{3}}} e^{-\frac{S}{3b}}. 
\end{eqnarray}
In order to check the validity of this approximation, numerical estimations of $K_YW$ are shown at Tables \ref{tab:infparam1} and \ref{tab:infparam2}. 

\section{Inflationary Cosmology and Inflationary Trajectory}

Because inflation is concerned with Planck scale physics, the dilaton can be one of the strong candidates to be identified with the inflaton\cite{03, 04}.
We will show here only the results from case 1 among the seven parameter choices for $\alpha$, $\beta$, for which the potential $V(S,Y)$ at $T=1$ has a stable minimum. 
Results for the other cases are summarized at Tables \ref{tab:infparam1} and \ref{tab:infparam2}. \\
First we show the result with the parameter choice, \quad $\beta = 6 \times 10^{-5}$, $c=10^2$ and $\alpha = 10^{-6}$ (Case 1).  Hereafter we fix 
$T=1$ ($\eta (1) = 0.768225$, $\eta' (1) = -0.192056$, $\eta'' (1) = -0.00925929$) and $b=\frac{15}{16\pi^2}$ corresponding to the $E_8$ gauge group. 
The minimum of the potential is given by
\begin{eqnarray}
S_{{\rm min}} = 2.23 \times 10^{-2}, \quad  
Y_{{\rm min}} = 1.12 \times 10^{-2}, \quad  
V(S_{{\rm min}},Y_{{\rm min}}) = 5.94 \times 10^{-12}.
\end{eqnarray}
The parameters of inflation are predicted as follows:
\begin{eqnarray}
&&S_{{\rm end}} = 0.7394, \quad S_* = 10.90, \quad \mathcal{P_{R^*}} = 2.438 \times 10^{-9}, \nonumber \\
&&N = 58.79, \quad\,\,\, n_{S^*} = 0.9746, \quad \alpha_{S^*} = -4.303 \times 10^{-4}.
\end{eqnarray}
The gravitino mass $M_{3/2}$ and and SUSY breaking scale $F_S$ are predicted as:
\begin{eqnarray}
M_{3/2} = | M_P \,\, e^{\frac{K}{2}} W | = 8.99 \times 10^{12} \,\, {\rm GeV}, \qquad 
F_S = 2.19 \times 10^{12} \,\, {\rm GeV}.
\end{eqnarray}
\newpage
\begin{figure}[!htbp]
\begin{minipage}{0.49\hsize}
We show the potential $V(S)$ minimized with respect to $Y$ in Fig. 1, the evolution of the slow-roll parameters in Fig. 2, and the stability of the potential minimum at $T=1$ in Fig. 3.\\
This case seems to explain the WMAP observations well.
The slow-roll parameters (in Planck units $M_{\rm P}/\sqrt{8\pi}=1$) are defined by
\begin{eqnarray} 
\epsilon_\alpha=\frac{1}{2}\left(\frac{\partial_\alpha V}{V}\right)^2, \qquad
\eta_{\alpha\beta}=\frac{\partial_\alpha\partial_\beta V}{V}.
\end{eqnarray}
The slow-roll condition is well satisfied, and the $\eta$-problem can be resolved. 
Using the SRA, the number of $e$-folds at which a comoving scale $k$ crosses the Hubble scale $aH$ during inflation is given by:
$N$ is also calculated by:
\begin{eqnarray}
N\sim -\int^{S_2}_{S_1}\frac{V}{\partial V}dS \sim 58.72,
\end{eqnarray}
by integrating from $S_{\rm end}$ to $S_*$, fixing the parameters $c\ {\rm and}\ b$ as well as $\alpha\ {\rm and}\ \beta$. That is, our potential has the ability to produce the cosmologically plausible number of $e$-folds.
The scalar spectral index $n_{s}$ for the scale dependence of the spectrum of density perturbation, and its tilt $\alpha_{s}$are estimated by substituting $S_*$ into these formula:
\begin{eqnarray} 
&&n_{s*}=1+\frac{d\ln \mathcal{P_R}}{d\ln k}\sim 0.9746, \\[5pt]
&&\alpha_{s*}=\frac{dn_s}{d\ln k}\sim -4.314 \times 10^{-4}. 
\end{eqnarray}
Finally, estimating the spectrum of the density perturbation caused by the slow-rolling dilaton, we obtain:
\begin{eqnarray}
\mathcal{P_R}\sim\frac{1}{12\pi^2}\frac{V^3}{\partial V^2}\sim 2.438 \times 10^{-9}.
\end{eqnarray}
The tensor power spectrum $\mathcal{P_T}$ and its ratio $r$ to $\mathcal{P_R}$ are estimated as
\begin{eqnarray}
&&\mathcal{P_{T^*}} = 1.652 \times 10^{-10}, \\[5pt]
&&r=\left( \frac{\mathcal{P_{T^*}}}{\mathcal{P_{R^*}}} \right) = 6.755 \times 10^{-2}.
\end{eqnarray}
\end{minipage}
\begin{minipage}{0.02\hsize}
\begin{center}
\end{center}
\end{minipage}
\begin{minipage}{0.49\hsize}
\begin{center}
\includegraphics[scale=0.8]{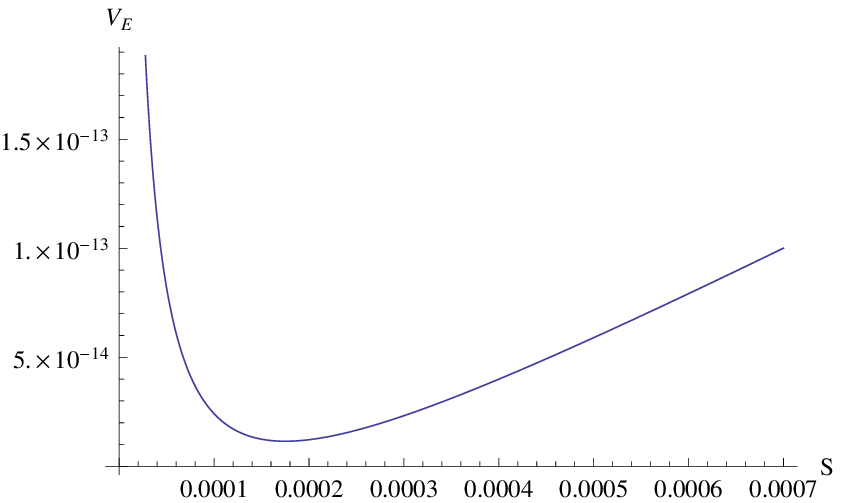}
\end{center}
\caption{\small The potential $V(S)$ minimized with \,\, respect to $Y$.}
$\,$ \\
\begin{center}
\includegraphics[scale=0.6]{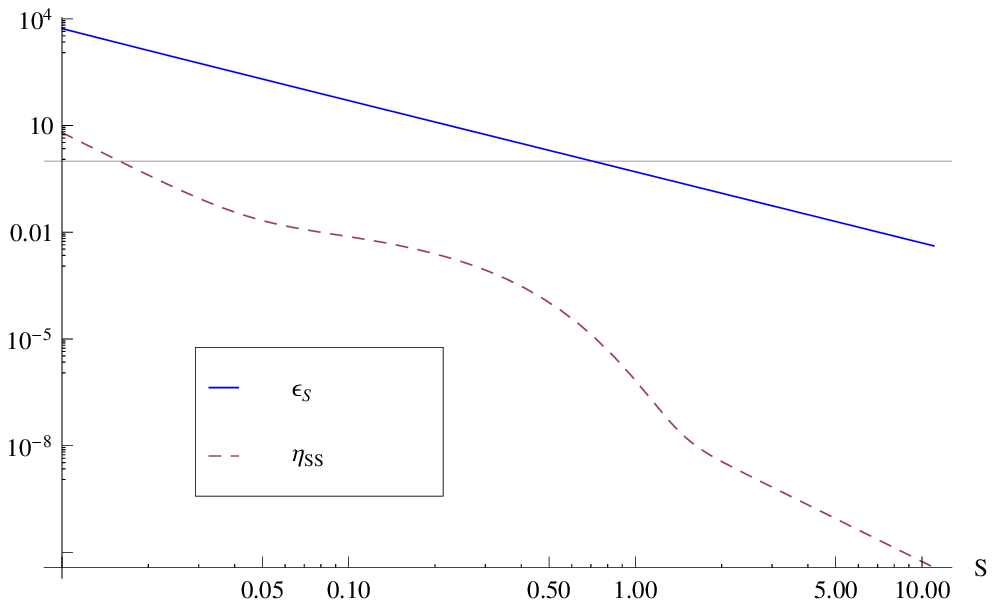}
\end{center}
\caption{\small The evolution of the slow-roll parameters. The blue curve represents $\epsilon_S$, while the red dashed curve denotes $|\eta_{SS}|$. }
$\,$ \\
\begin{center}
\includegraphics[scale=0.8]{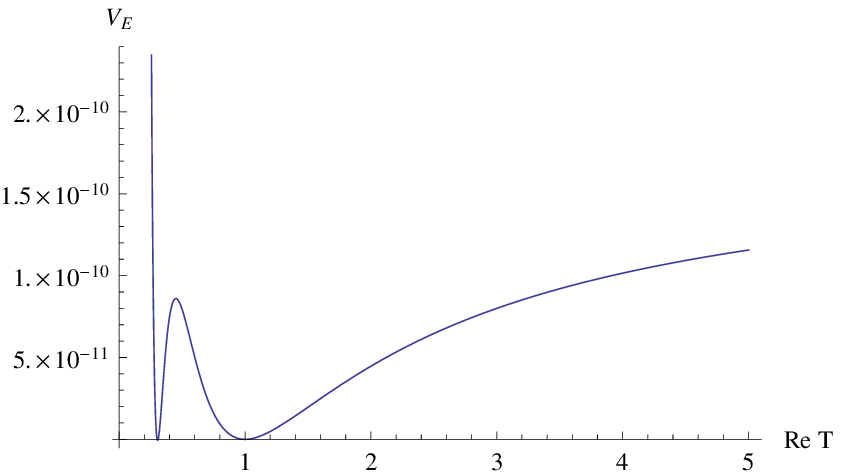}
\end{center}
\caption{\small The stability of the potential minimum at $T=1$.}
\end{minipage}
\end{figure}
$\,$ \\[-10pt]
For the potential at the minimum for the case 1, moreover, the energy scale is  
$V\sim10^{-12}$, which is non-negative. 
The results for the remaining six cases are almost identical to case 1, except the supergravity proper quantities, which are shown in Tables \ref{tab:infparam1} and \ref{tab:infparam2}.
It is the end of inflation, when the slow-roll parameter $\epsilon_\alpha$ or $\eta_{\alpha\beta}$ reaches the value 1. After passing through the minimum of the potential, reheating will begin. 

\section{Super Higgs Mechanism and Gravitino Production}

Let us consider the Super Higgs mechanism in our model.
The inflatino field $\tilde{S}$ with mass $m_{\tilde{S}}=0\ {\rm{GeV}}$, which is the SUSY partner of the inflaton (dilaton) field $S$, can play the role of the Higgsino field. Because the metric elements satisfy $g_{ST}=g_{SY}=0$ in the K\"{a}hler metric $g_{ij}$, $S$ does not mix with $Y,\ T$. Then the terms that cause the super Higgs mechanism (SHM) are selected as 
\begin{eqnarray}
{\mathcal{L}}_{\rm{SHM}} &=& e e^{\frac{G}{2}}\Big\{\psi_\mu\sigma^{\mu\nu}\psi_\nu+\bar{\psi}_\mu\bar{\sigma}^{\mu\nu}\bar{\psi}_\nu +\frac{i}{\sqrt{2}}G_S\tilde{S}\sigma^\mu\bar{\psi}_\mu+\frac{i}{\sqrt{2}}G_{S^*}\bar{\tilde{S}}\bar{\sigma}^\mu\psi_\mu \nonumber \\
&&+ \frac{1}{2}(G_{SS}+G_{S}G_{S})\tilde{S}\tilde{S}+\frac{1}{2}(G_{S^*S^*}+G_{S^*}G_{S^*})\bar{\tilde{S}}\bar{\tilde{S}}\Big\}. \\ \nonumber
\end{eqnarray}
First, we calculate the first, the fourth and the sixth terms as follows:
\begin{eqnarray}
{\mathcal{L}}_{\rm{SHM1}}
&=&ee^{\frac{G}{2}}\Big\{\Big(\psi_\mu+\frac{i}{3\sqrt{2}}G_{S^*}\bar{\tilde{S}}\bar{\sigma}_\mu\Big)\sigma^{\mu\nu}\Big(\psi_\nu-\frac{i}{3\sqrt{2}}G_{S^*}\sigma_\nu \bar{\tilde{S}}\Big) \nonumber \\
&&+ \frac{1}{2}(G_{S^*S^*}+\frac{1}{3}G_{S^*}G_{S^*})\bar{\tilde{S}}\bar{\tilde{S}}\Big\}\label{shm}.
\end{eqnarray}
Now the last term of eq.(\ref{shm}) implies the mass of $\tilde{S}$, which is proved to be exactly zero in our model. 
The first term can be identified with the mass term of the massive gravitino field, whose mass is given by $M_{3/2}=e^{G/2}$. 
The second, the third and the fifth terms are the Hermite conjugates of eq.(\ref{shm}), and similar expressions can be derived.
This is the scenario of the Super Higgs mechanism.  
The predicted values of the gravitino mass $M^{3/2}$ and the scale of SUSY breaking\cite{ref:Kallosh} 
\begin{align}
{F_S}^2 =  M_P^2 \, m_S g_{SS}^{-1} \, m_S + \frac{1}{2} M_P^2 \, \dot{S}^2
\end{align}
 in our model are summarized in Table \ref{tab:infparam1}.

Now we will investigate the gravitino production from heavy scalar bosons after inflation.
The interaction terms between the scalar fields $\phi_i$ and the gravitino $\psi_\mu$ in the total Lagrangian density of supergravity are selected as follows \cite{Moroi:1995fs}:
\begin{eqnarray}
&&e^{-1}{\cal{L}}_{int}=\epsilon^{\mu\nu\rho\sigma}\bar{\psi}_\mu\bar{\sigma}_{\nu}\partial_\rho\psi_\sigma
+\frac{1}{4}\epsilon^{\mu\nu\rho\sigma}\bar{\psi}_\mu\bar{\sigma}_{\nu}\left(G_j\partial_\rho\phi^j-G_{j^*}\partial_\rho\phi^{*j} \right)\psi_\sigma\nonumber\\
&&\qquad\qquad\quad -e^{G/2}\left(\psi_\mu\sigma^{\mu\nu}\psi_\nu+\bar{\psi}_\mu\bar{\sigma}^{\mu\nu}\bar{\psi}_\nu\right).
\end{eqnarray}
These interaction terms are expanded in terms of the shifts $\delta \phi_i$ from the stable values for each of the $\phi_i$'s, i.e., $\delta\phi_i=\phi_i-\left<\phi_i\right>$.
In our model there are three scalar fields $S,Y,T$ corresponding to the $\phi_i$'s.
$S,Y,T$ are canonically normalized by ${\phi_i}'=\sum_j\alpha^i_j\tilde{\phi^j}$, where $\alpha^i_j$ are
the coefficients of canonical normalization.
Since $\left<G_i\right>$'s are also affected by the normalization, the normalized $\left<{G_i}\right>$'s are replaced with $\left<G_i'\right>$s, and the 
$\alpha^i_j$'s are replaced by the $\left<{\alpha^i_j}\right>$'s at the stable points $\left<\tilde{\phi_i}\right>$.
By using these formula and the general relation for the gravitino mass $M_{3/2}=\left<e^{G/2}\right>$,
the interaction terms are obtained as:
\begin{eqnarray}
-\frac{1}{8\sqrt{2}} M_{3/2} \left<{G_i}' \right>\left<{\alpha^i_j}\right>\tilde{\phi^j}\bar{\psi}_\mu[\gamma^{\mu},\gamma^\nu]\psi_\nu.
\end{eqnarray}
The helicity $1/2$ part of the massive gravitino is defined by the tensor product of a vector and a spinor as \cite{Auvil:1966} 
\begin{eqnarray}
u_\mu(k;1/2)\simeq i\sqrt{\frac{2}{3}}\epsilon_\mu(k;0) u(k;1/2)+i\sqrt{\frac{1}{3}}\epsilon_\mu(k;1)u(k;-1/2),
\end{eqnarray}
where the coefficient of each term is a Clebsch-Gordan coefficient, $\epsilon_\mu(k;\lambda)$ is the wave function of vector field with helicity $\lambda$, and $u(k;h)$ is the spinor wave function with helicity $h$.
The decay rate $d\Gamma=\frac{|{\bf{k}}|}{8\pi m_\phi^2}|{\cal{M}}|^2\frac{d\Omega}{4\pi}$ is obtained as \cite{Endo:2006zj,Endo:2006xg,Endo:2007sz, Nakamura:2006uc,Asaka:2006bv}:
\begin{eqnarray}
\Gamma(\phi\rightarrow \psi_{3/2}+\psi_{3/2})=\frac{\left<{G_i}' \right>^2\left<{\alpha^i_j}\right>^2 M_\phi^3}{288\pi}\left(1-\frac{2 M^2_{3/2}}{M_\phi ^2}\right)^2\left(1-\frac{4 M^2_{3/2}}{M_\phi ^2}\right)^\frac{1}{2}\label{decayrate2}.
\end{eqnarray}
After the scalars $S,Y,T$ are canonically normalized and the masses diagonalized, the mass eigenstates are denoted by $S'',Y'',T''$, 
and the masses are calculated as 
\begin{eqnarray}
M_{S''} &=& 9.982 \times 10^{12} {\rm GeV}, \qquad\qquad M_{Y''}=2.606 \times 10^{16} {\rm GeV}, \nonumber \\[5pt]
M_{T''} &=& 2.265 \times 10^{12} {\rm GeV},
\end{eqnarray}
for the case 1. Therefore, we obtain for the
decay process $Y'' \rightarrow \psi_{3/2}+\psi_{3/2}$, by using the above formula:
\begin{eqnarray}
\Gamma(Y'' \rightarrow \psi_{3/2}+\psi_{3/2}) &=& 4.785 \times 10^4 \,\, {\rm GeV}, \\[5pt]
\tau (Y'' \rightarrow \psi_{3/2}+\psi_{3/2})  &=& 1.376 \times 10^{-29} \,\, {\rm sec}.
\end{eqnarray}
This process occurs almost instantly. For the other cases, see Tables \ref{tab:infparam1} and \ref{tab:infparam2}. 

\section{Reheating Temperature}

As an example, the decay rate of $S''$ into gauginos is estimated in our model. 
By using the term 
${\mathcal{L}}_{gaugino}=\kappa \int d^2\theta f_{ab}(\phi)W_{\alpha}W^{\alpha}, \\ f_{ab}(\phi)=\phi\delta_{ab}$, 
the interaction between $S$ and the gauginos $\lambda^a$'s are given as
\begin{eqnarray}
&&{\mathcal{L}}_{gaugino}
=\frac{i}{2}f^R_{ab}(\phi)\left[\lambda^a\sigma^\mu\tilde{\mathcal{D}}_\mu\bar{\lambda}^b+\bar{\lambda}^a\sigma^\mu\tilde{\mathcal{D}}_\mu\lambda^b \right]-\frac{1}{2}f^I_{ab}(\phi)\tilde{\mathcal{D}}_\mu\left[\lambda^a\sigma^{\mu}\bar{\lambda}^b\right] \nonumber \\
&&\qquad -\frac{1}{4}\frac{\partial f_{ab}(\phi)}{\partial \phi}e^{K/2}G_{\phi\phi^*}D_{\phi^*}W^*\lambda^a\lambda^b +\frac{1}{4}\left(\frac{\partial f_{ab}(\phi)}{\partial \phi}\right)^*e^{K/2}G_{\phi\phi^*}D_{\phi}W\bar{\lambda}^a\bar{\lambda}^b. \label{gaugino_decay}
\end{eqnarray}
By looking at the first term of (\ref{gaugino_decay}), the $\lambda^a$'s are also canonically normalized as $\lambda^a= \left<f^R_{ab}\right>^{-\frac{1}{2}}\hat{\lambda}^a$. 
The interactions come from the third and fourth terms. The terms include $e^{K/2}G^{\phi\phi^*}D_{\phi^*}W^*$, which implies the auxiliary field of $\phi$ in global SUSY theory and it is replace by 
$F_\phi$.\\
By expanding $\frac{\partial f_{ab}}{\partial\phi}F_\phi$ around the stable point, the interaction terms are given as
\begin{eqnarray}
\!\!\!\!\!\! {\mathcal{L}}_{\rm{int}}  
=-\frac{1}{4\left<f_{ab}\right>}\left[\left<\frac{\partial^2 f_{ab}}{\partial\phi^2}F_\phi+\frac{\partial f_{ab}}{\partial\phi}\frac{\partial F_\phi}{\partial\phi}\right>\delta\phi+\left<\frac{\partial f_{ab}}{\partial\phi}
\frac{\partial F_\phi}{\partial\phi^*}\right>\delta\phi^* \right]\lambda^a\lambda^b + {\rm h.c.} .
\end{eqnarray}
When $\phi = S$, $F_S$ implies the SSB scale of the model and is estimated as $\left< S+S^* \right> \gg M_{3/2}$, since $\left< F_S \right> \sim M_{3/2}$ and $(S+S^*)$ take values close to the 
Planck scale. Therefore, as the first term is far smaller than the second and actually negligible, $- \left< \frac{\partial F_S}{\partial S} \right> \sim M_{3/2}$ remains.
The derivative term of $S^*$ can be replaced by $- \left< \frac{\partial F_S}{\partial S^*} \right> \sim m_{S}$. Then the decay rate $\Gamma(\phi\to \lambda+\lambda)$ can be estimated as: 
\begin{eqnarray}
\Gamma(S'' \to \lambda + \lambda)=\frac{3}{16\pi}\frac{\left<\alpha^i_j\right>^2}{\left<f_{ab}\right>^2} M^2_\lambda M_{S''} \left(1+\frac{M^2_{3/2}}{M^2_{S''}}+2\frac{M_{3/2}}{M_{S''}}\right)\left(1-\frac{4M_\lambda^2}{M_{S''}^2} \right)^\frac{1}{2}.
\end{eqnarray}
The reheating temperature $T_R({\rm gaugino})$ is derived from the Boltzmann equation by using the decay rate, and is given by
\begin{eqnarray}
T_R({\rm gaugino})=\left(\frac{10}{g_*}\right)^\frac{1}{4}\sqrt{M_P~\Gamma(S'' \to \lambda + \lambda) },
\end{eqnarray}
where $g_*$ is the number of effective degrees of freedom of MSSM, i.e.  $g_* =228.75$ and numerically given above by inserting the decay rate from the canonically normalized inflaton field $S''$. 
\\
By using the relation $F_S \sim M_P \, M_{SP}$, which holds for the mass of SUSY particles (SP)\cite{ref:Polchinski},
the gaugino masses are estimated as \\
\begin{equation}
M_\lambda = \frac{F_S^2}{M_P} = 1.971 \times 10^6 \,\, {\rm GeV}. 
\end{equation}
Then the decay rate of  $S'' \rightarrow \lambda+\lambda$  and the reheating temperature are estimated as
\begin{eqnarray}
\Gamma(S'' \to \lambda+\lambda) = 2.96 \times 10^{-3}  \,\, {\rm GeV}, \quad \quad
T_R({\rm gaugino}) = 3.880 \times 10^{7} \,\, {\rm GeV} . 
\end{eqnarray}
Because the primordial gravitinos decay very rapidly and the reheating temperature is lower than the gravitino mass,  gravitino reproduction will not occur after reheating. The the effect on the standard Big Bang Nucleosynthesis (BBN) scenario \cite{Endo:2007ih,Takahashi:2007tz, Kawasaki:1994af, Kawasaki:2004qu, Kawasaki:2008qe,Kawasaki:2006gs,Kawasaki:2006hm, Giudice:1999am, Bolz:2000fu, Khlopov:1999}) may be negligible in our model (see also for the bound of reheating temperature\cite{Weinberg:1982zq, Khlopov:1984pf}).

All these numerical values are shown in Tables \ref{tab:infparam1} and \ref{tab:infparam2}, and will be explained in next Section.

\section{Numerical predictions by tuning parameters $\beta,\ \alpha$ and $c$}

The numerical predictions of inflationary variables coming from the scalar potential in SRA are shown in Tables \ref{tab:infparam1} and \ref{tab:infparam2}. 
Those values are extremely well fitted to the WMAP observations for all the Cases 1 to 7.
Here it will be instructive to revisit our scenario, which should be investigated by experiments.
The renormalization group $\beta (g)$ function is assumed to be fixed, corresponding to the $E_8$ hidden sector gauge group, i.e.,
\begin{align}
b &= \frac{\beta_0}{96 \pi^2}= \frac{15}{16 \pi^2},
\end{align}
because the $\beta$ function is defined as
$\beta (g) = - \frac{\beta_0}{16 \pi^2} g^3
= - \frac{g^3}{4 \pi^2} \left( \frac{9}{11} \frac{11}{12} C_2 \right) = - \frac{g^3}{4 \pi^2} \frac{3}{4} C_2$,
Therefore, $\beta_0 = 3 C_2,\quad C_2 (E_8) = 30$.
Next we also fix the modular field at $T=1$, where the potential is required at least to be stable.
Varying the parameters $\beta,\ \alpha$ and $c$, we can see that the slow roll inflation parameters are almost completely compatible with 7-year WMAP observations in the seven cases we have chosen. In these numerical estimates of inflation, the parameter $\beta$ is intimately related to the value of the power spectrum $\mathcal{P_{R^*}}$. We find 
\begin{align}
\beta &= 6 \times 10^{-5},
\end{align}
which realizes the slow roll inflation to explain the WMAP observations almost exactly. The other two parameters $\alpha$ and $c$ are essentially playing the roles to determine the prediction of the SUSY part of the theory, such as gravitino and gaugino masses. The parameter $\alpha$ is fixed to 
\begin{align}
\alpha &= 10^{-8},
\end{align}
from case 2 to 7. 
The inflationary variables $N_{S^*}$, $n_{S^*}$, $\alpha_{S^*}$, and $\mathcal{P_{R^*}}$ are then calculated. All the values fit extremely well to the WMAP observations. The values of $\mathcal{P_{T^*}}$ and $r$ are new predictions of our present model, in which the $r$'s appear to be about $6.8 \times 10^{-2}$ in all cases. We expect that the Planck satellite will prove these predictions soon. We estimated the values of variables of supergravity, i.e., the gravitino mass $M^{3/2} \,\, ({\rm GeV})$, the SUSY breaking scale $F_S \,\, ({\rm GeV})$, the order of the gaugino masses $M_\lambda \,\, ({\rm GeV})$, and the reheating temperature $T_R({\rm gaugino}) \,\, ({\rm GeV})$, as well as the decay rates of gravitino and heavy scalar $Y$.     
The dependences on the parameter $c$ of the SUSY sector can be seen in the Tables, among which some cases show the possibility of gauginos to be observed by LHC or other experiments aiming to search dark matter. The validity of our model may be clearly checked by the experiments.
The parameter dependence of our model will be explained more thoroughly in a separate paper.\\
We summarize our results for the seven cases in Tables \ref{tab:infparam1} and \ref{tab:infparam2}.

\setlength{\arrayrulewidth}{.3pt}
\setlength{\tabcolsep}{7pt}
\begin{table}[!htbp]
\caption{Predictions of inflation parameters and SUSY particles ($\beta = 6 \times 10^{-5}$, $b=\frac{15}{16 \pi^2}$)}
\label{tab:infparam1}
\begin{flushleft}
\begin{tabular}{cccccc}
\hline\hline
Parameters & Case 1 & Case 2 & Case 3 & Case 4  \\ \hline
$c$ & $100$ & $1000$ & $869$ & $769$  \\
$\alpha$ & $10^{-6}$ & $10^{-8}$ & $10^{-8}$ & $10^{-8}$ \\
$S_{min}$ & $2.234 \times 10^{-2}$ & $1.722 \times 10^{-4}$ & $1.750 \times 10^{-4}$ & $1.785 \times 10^{-4}$  \\
$Y_{min}$ & $1.123 \times 10^{-2}$ & $1.213 \times 10^{-3}$ & $1.396 \times 10^{-3}$ & $1.578 \times 10^{-3}$  \\
$V(S_{{\rm min}},Y_{{\rm min}})$ & $5.938 \times 10^{-12}$ & $7.574 \times 10^{-15}$ & $1.149 \times 10^{-14}$ & $1.650 \times 10^{-14}$ \\
$K_YW$ & $3.715 \times 10^{-8}$ & $3.670 \times 10^{-11}$ & $4.813 \times 10^{-11}$ & $4.813 \times 10^{-11}$ \\
$S_{{\rm end}}$ & $0.7394$ & $0.7074$ & $0.7074$ & $0.7074$ \\
$S^*$ & $10.90$ & $10.86$ & $10.86$ & $10.86$ \\
$N_{S^*}$ & $58.72$ & $58.72$ & $58.72$ & $58.72$ \\
$n_{S^*}$ & $0.9746$ & $0.9746$ & $0.9746$ & $0.9746$ \\
$\alpha_{S^*}$ & $-4.314 \times 10^{-4}$ & $-4.314 \times 10^{-4}$ & $-4.314 \times 10^{-4}$ & $-4.314 \times 10^{-4}$ \\
$\mathcal{P_{R^*}}$ & $2.438 \times 10^{-9}$ & $2.433 \times 10^{-9}$ & $2.433 \times 10^{-9}$ & $2.433 \times 10^{-9}$ \\
$\mathcal{P_{T^*}}$ & $1.652 \times 10^{-10}$ & $1.650 \times 10^{-10}$ & $1.650 \times 10^{-10}$ & $1.650 \times 10^{-10}$ \\
$r$ & $6.755 \times 10^{-2}$ & $6.784 \times 10^{-2}$ & $6.784 \times 10^{-2}$ & $6.784 \times 10^{-2}$ \\
$M^{3/2} \,\, ({\rm GeV})$ & $8.987 \times 10^{12}$ & $9.355 \times 10^{11}$ & $9.316 \times 10^{11}$ & $9.268 \times 10^{11}$ \\
$F_S \,\, ({\rm GeV})$ & $2.191 \times 10^{12}$ & $2.332 \times 10^{10}$ & $3.495 \times 10^{10}$ & $4.942 \times 10^{10}$ \\
$M_\lambda \,\, ({\rm GeV})$ & $1.971 \times 10^6$ & $223.2$ & $501.6$ & $1003$ \\
\hline
\end{tabular}
\end{flushleft}
$\,$ \\
\begin{flushleft}
\begin{tabular}{cccccccc}
\hline\hline
Parameters & Case 5 & Case 6 & Case 7 \\ \hline
$c$ & $571$ & $498$ & $349$  \\
$\alpha$ & $10^{-8}$ & $10^{-8}$ & $10^{-8}$  \\
$S_{min}$ & $1.936 \times 10^{-4}$ & $2.051 \times 10^{-4}$ & $2.520 \times 10^{-4}$  \\
$Y_{min}$ & $2.125 \times 10^{-3}$ & $2.436 \times 10^{-3}$ & $3.476 \times 10^{-3}$  \\
$V(S_{{\rm min}},Y_{{\rm min}})$ & $3.942 \times 10^{-14}$ & $5.843 \times 10^{-14}$ & $1.579 \times 10^{-13}$  \\
$K_YW$ & $1.102 \times 10^{-10}$ & $7.649 \times 10^{-11}$ & $1.102 \times 10^{-10}$ \\
$S_{{\rm end}}$ & $0.7074$ & $0.7074$ & $0.7074$  \\
$S^*$ & $10.86$ & $10.86$ & $10.86$  \\
$N_{S^*}$ & $58.72$ & $58.72$ & $58.72$ \\
$n_{S^*}$ & $0.9746$ & $0.9746$ & $0.9746$ \\
$\alpha_{S^*}$ & $-4.314 \times 10^{-4}$ & $-4.314 \times 10^{-4}$ & $-4.314 \times 10^{-4}$ \\
$\mathcal{P_{R^*}}$ & $2.433 \times 10^{-9}$ & $2.433 \times 10^{-9}$ & $2.433 \times 10^{-9}$ \\
$\mathcal{P_{T^*}}$ & $1.650 \times 10^{-10}$ & $1.650 \times 10^{-10}$ & $1.650 \times 10^{-10}$ \\
$r$ & $6.784 \times 10^{-2}$ & $6.784 \times 10^{-2}$ & $6.784 \times 10^{-2}$ \\
$M^{3/2} \,\, ({\rm GeV})$ & $9.060 \times 10^{11}$ & $8.899 \times 10^{11}$ & $8.105 \times 10^{11}$ \\
$F_S \,\, ({\rm GeV})$ & $1.108 \times 10^{11}$ & $1.566 \times 10^{11}$ & $3.502 \times 10^{11}$ \\
$M_\lambda \,\, ({\rm GeV})$ & $5038$ & $1008 \times 10$ & $5035 \times 10$ \\
\hline
\end{tabular}
\end{flushleft}
\end{table}
\newpage

\setlength{\arrayrulewidth}{.3pt}
\setlength{\tabcolsep}{7pt}
\begin{table}[!htbp]
\caption{Predictions of inflation parameters and SUSY particles}
\label{tab:infparam2}
\begin{flushleft}
\begin{tabular}{cccccc}
\hline\hline
Parameters & Case 1 & Case 2 & Case 3 & Case 4  \\[3pt] \hline
$M_{S''} \,\, ({\rm GeV})$ & $9.982 \times 10^{12}$ & $9.366 \times 10^{11}$ & $9.342 \times 10^{11}$ & $9.320 \times 10^{11}$ \\
$M_{Y''} \,\, ({\rm GeV})$ & $2.606 \times 10^{16}$ & $3.208 \times 10^{16}$ & $3.662 \times 10^{16}$ & $4.098 \times 10^{16}$ \\
$M_{T''} \,\, ({\rm GeV})$ & $2.265 \times 10^{12}$ & $2.332 \times 10^{10}$ & $3.495 \times 10^{10}$ & $4.942 \times 10^{10}$ \\
$M_{\tilde{Y}} \,\, ({\rm GeV})$ & $1.541 \times 10^{15}$ & $2.050 \times 10^{14}$ & $2.693 \times 10^{14}$ & $3.405 \times 10^{14}$ \\
$M_{\tilde{T}} \,\, ({\rm GeV})$ & $2.198 \times 10^{13}$ & $1.984 \times 10^{13}$ & $2.425 \times 10^{13}$ & $3.274 \times 10^{13}$ \\
$\Gamma(Y'' \to 2\psi_{3/2}) \,\, ({\rm GeV})$ & $4.785 \times 10^{4}$ & $7.572 \times 10^{13}$ & $2.133 \times 10^{4}$ & $3.441 \times 10^{4}$ \\
$\Gamma(S'' \to 2\lambda) \,\, ({\rm GeV})$ & $2.957 \times 10^{-3}$ & $2.423 \times 10^{-12}$ & $5.588 \times 10^{-12}$ & $1.151 \times 10^{-11}$ \\
$T_R({\rm gaugino}) \,\, ({\rm GeV})$ & $3.880 \times 10^{7}$ & $1.111 \times 10^{3}$ & $1.687 \times 10^{3}$ & $2.421 \times 10^{3}$ \\
\hline
\end{tabular}
\end{flushleft}
$\,$ \\
\begin{flushleft}
\begin{tabular}{cccccccc}
\hline\hline
Parameters & Case 5 & Case 6 & Case 7 \\ \hline
$M_{S''} \,\, ({\rm GeV})$ & $9.327 \times 10^{11}$ & $9.434 \times 10^{11}$ & $1.107 \times 10^{12}$ \\
$M_{Y''} \,\, ({\rm GeV})$ & $5.299 \times 10^{16}$ & $5.903 \times 10^{16}$ & $7.597 \times 10^{16}$ \\
$M_{T''} \,\, ({\rm GeV})$ & $1.108 \times 10^{11}$ & $1.566 \times 10^{11}$ & $3.502 \times 10^{11}$ \\
$M_{\tilde{Y}} \,\, ({\rm GeV})$ & $5.931 \times 10^{14}$ & $7.577 \times 10^{14}$ & $1.392 \times 10^{14}$ \\
$M_{\tilde{T}} \,\, ({\rm GeV})$ & $6.044 \times 10^{13}$ & $7.324 \times 10^{13}$ & $6.616 \times 10^{13}$ \\
$\Gamma(Y'' \to 2\psi_{3/2}) \,\, ({\rm GeV})$ & $8.103 \times 10^{4}$ & $9.122 \times 10^{4}$ & $7.933 \times 10$ \\
$\Gamma(S'' \to 2\lambda) \,\, ({\rm GeV})$ & $6.423 \times 10^{-11}$ & $1.354 \times 10^{-10}$ & $6.641 \times 10^{-10}$ \\
$T_R({\rm gaugino}) \,\, ({\rm GeV})$ & $5.719 \times 10^{3}$ & $8.303 \times 10^{3}$ & $1.839 \times 10^{4}$ \\
\hline
\end{tabular}
\end{flushleft}
\end{table}
$\,$ \\[-40pt]

\section{Summary}

By modifying the original string-inspired modular invariant supergravity, which are known to explain WMAP observations appropriately, a mechanism of SSB and Gravitino production just after the end of inflation as been investigated.
The model we used cleared the $\eta$-problem and the negative energy problem of the potential at the stable point, and appeared to reproduce successfully the observations on the inflation era. The assumption to set $W_Y=0$ instead of imposing $W_Y+K_YW=0$ in order to determine the minimum of the scalar potential, which seems to be a rather good approximation, might give rise to some problems. However, we were able to check that $F_Y$ itself has a very small SUSY breaking contribution and therefore is no flat direction.     
Seven cases of parameter choices have been discussed here, among which some examples (Cases 2 to 5) show the possibility of gauginos to be observed by LHC experiments, which will give some hints on the identity of dark matters. We would like to emphasize that the ratio $r\sim 6.8 \times 10^{-2}$ between the scalar power spectrum and the tensor one is predicted to be almost the same for all seven cases, and this value seems to be in the range possibly observed by the Planck satellite soon. 
 \\
 Because the supergravity seems the most plausible framework to explain new physics, including undetected objects, the supergravity model of inflation which here described will open the hope to construct a realistic theory of particles and cosmology in this framework.

\end{document}